\documentstyle[preprint,aps]{revtex}
\begin{document}
\draft
\title{
Antiferromagnetic symmetry breaking in the\\
half filled Hubbard model in infinite dimensions}
\author{Peter Kopietz}
\address{
Institut f\H{u}r Theoretische Physik der Universit\H{a}t G\H{o}ttingen,\\
Bunsenstr.9, D-3400 G\H{o}ttingen, Germany}
\date{\today}
\maketitle
\begin{abstract}
We study the half filled Hubbard model on a hypercubic lattice
in infinite dimensions in the presence of a staggered magnetic field.
An exact Ward-identity between vertex functions
and self-energies is derived, that holds in any phase
without broken symmetry for all values of $U$.
Making the reasonable assumptions that for small enough on-site
repulsion $U$ the high-temperature phase is a Fermi liquid, and that
in the weak coupling regime the effective Anderson
impurity model can be studied perturbatively, we proof that Hartree-Fock theory
and the random-phase approximation are very accurate for small $U$, and that
the system develops long-range antiferromagnetic order at a finite temperature.

\end{abstract}
\narrowtext
%
%
\section{Introduction}
\setcounter{section}{1}
\label{sec:intro}
The physics of the Hubbard model in $d= \infty$ dimensions
is currently investigated by several
groups\cite{Rozenberg92}-\cite{Bartkowiak93}.
The essential simplification in infinite dimensions is that the
self-energy $\Sigma ( {\bf{k}} , \omega )$ is independent of
the wave vector ${\bf{k}}$\cite{MullerHartmann89}.
At the same time, the dependence on the frequency $\omega$ remains non-trivial,
and reflects generic features of correlated electronic systems.
Recent works have mainly focused on Fermi liquid- and Mott insulating
phases\cite{Rozenberg92,Georges92a}.
Although antiferromagnetic symmetry breaking has been briefly discussed in the
literature\cite{Georges92c,Vollhardt92}, and has been studied numerically in
Ref.\cite{Jarrell92},
it seems that the functional-integral equations
describing phases with broken symmetry have not been
investigated analytically.
A perturbative study of antiferromagnetism in infinite dimensions
has recently been published in Ref.\cite{Bartkowiak93}.
However, within this approach one finds at weak coupling a Neel temperature
$T_N$ that is
almost linear in the on-site repulsion $U$, in contradiction with
Hartree-Fock theory, which predicts that $T_N$ is exponentially small.
In the present work we shall use a non-perturbative method to study
antiferromagnetic symmetry
breaking in the repulsive half filled Hubbard model in infinite dimensions.
We shall mainly focus on the weak coupling regime, and show that
in this regime
Hartree-Fock theory is essentially exact.

It is generally accepted that, at least in $d \geq 3$, the model
has for any finite positive $U$ antiferromagnetic
long-range order (LRO) at sufficiently low temperature.
However, a rigorous proof of LRO does not
exist. In particular, it is not obvious that the LRO is maintained
for arbitrarily small $U >0$,
because the mean-field result for the order parameter at weak coupling
is exponentially small, and fluctuations may be important.
Here we shall consider
Hubbard models with perfect nesting, particle-hole symmetry, and
finite  density of states at the Fermi energy.
We shall proof that
in $d = \infty$ there exists for sufficiently small but finite $U > 0$
a finite temperature $T_{N}(U) > 0$, such that the
susceptibility for the staggered magnetization is divergent for all
temperatures $T < T_{N} (U)$.
Provided that no other susceptibility diverges at
temperatures larger than $T_{N}$, our result implies
the existence of long-range antiferromagnetic order in the weak coupling regime
at low but finite temperatures.

One way to examine antiferromagnetic symmetry breaking
is to calculate the staggered magnetization ${\bf{M}} ( {\bf{h}} )$
in the presence of a
staggered field ${\bf{h}}$, take the thermodynamic limit, and then
let ${\bf{h}} \rightarrow 0$.
If the spin-rotational symmetry is spontaneously broken,
${\bf{M}}_{0} \equiv \lim_{\bf{h} \rightarrow 0 } {\bf{M}} ( {\bf{h}} ) $
does not vanish.
Alternatively, one can calculate the  staggered susceptibility
$\chi ( T  , U)$ in a parameter regime where ${\bf{M}}_{0} = 0$.
Antiferromagnetic symmetry breaking manifests itself in a divergence
of $\chi (T , U )$ at
the Neel temperature $T_{N} (U)$. In the present work we shall
take the latter approach and show that in infinite dimensions
perfect nesting and
particle-hole symmetry imply in the weak coupling regime that $T_{N} (U) > 0$
for $U > 0$.

The hamiltonian of the Hubbard models under consideration is given by
${\cal{H}} = {\cal{H}}_{0} + {\cal{U}}$, with
 \begin{eqnarray}
 {\cal{H}}_{0} & = & -
 \sum_{\bf{R} , \bf{r}  }
 {t_{ \bf{r} }}
 c^{\dagger}_{\bf{R}  } c_{\bf{R} + \bf{r}  }
 - {\bf{h}} \cdot
 \sum_{\bf{R}}
 c^{\dagger}_{\bf{R} } \vec{\sigma} c_{\bf{R}}
 e^{ i {\bf{\Pi}} \cdot {\bf{R}} }
 \label{eq:Hamiltonian0}
 \\
 {\cal{U}} & = &
 U \sum_{\bf{R}} \left[ c^{\dagger}_{\bf{R}  \uparrow} c_{\bf{R} \uparrow} -
\frac{1}{2} \right]
 \left[ c^{\dagger}_{\bf{R}  \downarrow} c_{\bf{R} \downarrow} - \frac{1}{2}
\right]
  \label{eq:Hamiltonian1}
 \; \; \; ,
 \end{eqnarray}
where the ${\bf{R}}$-sum is over $N$ sites of a $d$-dimensional hypercubic
lattice, ${\bf{h}}$ is a
staggered field, and ${\bf{\Pi}} = [ \pi , \ldots , \pi ]$ is the
antiferromagnetic ordering
vector (we set the lattice spacing equal to unity).
We have defined two-component operators
$c^{\dagger}_{\bf{R}} = [c_{\bf{R} \uparrow} , c_{\bf{R} \downarrow } ]$,
where $c^{\dagger}_{\bf{R} \sigma}$ creates spin-$\sigma$ fermions at site
${\bf{R}}$,
and $\vec{\sigma} = [ \sigma^{x} , \sigma^{y} , \sigma^{z} ]$ are the Pauli
matrices.
We allow only hoppings that connect
different sublattices, so that
$e^{ i \bf{\Pi} \cdot {\bf{r}} } = -1$. This implies that the band
structure of the non-interacting model, defined by
 \begin{equation}
 \epsilon_{\bf{k}} = - \sum_{\bf{r}}
 t_{\bf{r}}
 e^{i {\bf{k}} \cdot {\bf{r}} }
 \label{epsiondef}
 \end{equation}
satisfies the perfect nesting condition
 \begin{equation}
 \epsilon_{\bf{k + \Pi }} = - \epsilon_{\bf{k}}
 \; \; \; .
 \label{eq:nesting}
 \end{equation}
Metzner and Vollhardt\cite{Metzner89}
first pointed out that a non-trivial limit $d \rightarrow \infty$
is only obtained if the $t_{\bf{r}}$ are
properly rescaled with inverse powers of $d$ to compensate for the
increase in the number of neighbors in high dimensions.
The nearest neighbor hopping energy should be scaled as $t_{\bf{r}} = t/
\sqrt{2d}$.
For general hoppings connecting different sublattices,
we require that  $t_{\bf{r}}$ vanishes
for large $d$ in such a way that
for $d \rightarrow \infty$
the density of states at $U=0$ has a finite limit $\rho ( \epsilon )$,
 \begin{equation}
 \rho ( \epsilon ) = \lim_{d \rightarrow \infty }
 \lim_{N \rightarrow \infty} \frac{1}{N} \sum_{\bf{k}} \delta ( \epsilon -
\epsilon_{\bf{k}} )
 \; \; \; ,
 \label{eq:rhodef}
 \end{equation}
where the wave-vector sum is over the first Brillouin zone.
Note also that the hopping energies depend only in the distance between the
sites,
and do not break the translational invariance of the lattice.
This is sufficient to assure that\cite{Dongenpriv}
 \begin{equation}
 \rho (0) > 0
 \; \; \; .
 \label{eq:rholarger}
 \end{equation}
We have included the terms proportional to the density in the
definition of the interaction in Eq.\ref{eq:Hamiltonian1},
because then at half filling the chemical potential is exactly zero at any
temperature, and
the Hartree correction to the self-energy vanishes.
The spectrum of our model has then particle-hole symmetry
at half filling.
The purpose of this paper is to show that
this property, together with
Eqs.\ref{eq:nesting} and \ref{eq:rholarger},
are sufficient to imply that for $ \rho(0) U \ll 1$
our model model has long-range antiferromagnetic order
at sufficiently low but finite temperatures.
%
%
\section{Antiferromagnetism in infinite dimensions}
\setcounter{section}{2}
\setcounter{equation}{0}
\label{sec:sym}

In this section we shall derive an exact functional-integral equation
for the self-energies in the presence of a staggered field. Although
an equivalent equation has been written down in Ref.\cite{Georges92c},
we use here an unconventional basis that greatly simplifies the
following analysis.

Imposing the usual periodic boundary conditions, the
free part ${\cal{H}}_{0}$ of our hamiltonian can
be transformed into block-diagonal form via Fourier transformation,
 \begin{equation}
 c_{\bf{k} \sigma} = N^{-1/2} \sum_{\bf{R}} e^{i {\bf{k}} \cdot {\bf{R}}}
c_{\bf{R} \sigma}
 \label{eq:ft}
 \; \; \; .
 \end{equation}
Conventionally, one chooses the staggered field in the $z$-direction,
${\bf{h}} \cdot \vec{\sigma}  = h  \sigma^{z}$. In this case ${\cal{H}}_{0}$
can be written as
 \begin{equation}
 {\cal{H}}_{0}^{z} = \sum_{{\bf{k}} \in RBZ } \sum_{\sigma}
 \tilde{C}^{\dagger}_{\bf{k} \sigma }
 \left( \begin{array}{cc}
 \epsilon_{\bf{k}} & - \sigma h  \\
 - \sigma h  & \epsilon_{\bf{k} + \bf{\Pi}}
 \end{array} \right)
 \tilde{C}_{\bf{k} \sigma }
 \label{eq:Sz}
 \; \; \; ,
 \end{equation}
where the momentum sum is over the reduced Brillouin zone of the
antiferromagnet, and
$\tilde{C}^{\dagger}_{\bf{k} \sigma} =
[ c^{\dagger}_{\bf{k} \sigma}  , c^{\dagger}_{\bf{k} + {\bf{\Pi}} \sigma} ]$.
Note that the sign of the off-diagonal elements in the quadratic form in
Eq.\ref{eq:Sz}
depends on the spin projection. For a derivation of the functional
integral-equation for
the exact self-energy in $d = \infty$ this introduces unpleasant technical
difficulties,
because we have to deal with two-component operators that carry in addition a
spin index.
A simple trick to avoid this difficulty is
to choose the staggered field in the $x$-direction\cite{Georges91},
${\bf{h}} \cdot \vec{\sigma} = h \sigma^{x}$.
In this case Eq.\ref{eq:Hamiltonian0} can be written as
 \begin{equation}
 {\cal{H}}_{0}^{x} = \sum_{{\bf{k}}}
 {C}^{\dagger}_{\bf{k}  }
 \left( \begin{array}{cc}
 \epsilon_{\bf{k}} & - h\\
 -  h & \epsilon_{\bf{k} + \bf{\Pi}}
 \end{array} \right)
 {C}_{\bf{k}  }
 \label{eq:Sx}
 \; \; \; ,
 \end{equation}
where the two-component operators are now defined by
 \begin{equation}
 {C}^{\dagger}_{\bf{k} } =
 \left[
  c^{\dagger}_{\bf{k} \uparrow}  , c^{\dagger}_{\bf{k} + {\bf{\Pi}} \downarrow}
 \right] \; \; \; .
 \label{eq:Cdef}
 \end{equation}
Note that the $C_{\bf{k}}$ are composed from operators with different spin
projection,
and that the sum in Eq.\ref{eq:Sx} is over the full Brillouin zone.
Loosely speaking, the antiferromagnetic
symmetry breaking is now labeled by a spin flip, so that the extra spin
summation
in Eq.\ref{eq:Sz} can be absorbed in the second component of $C_{\bf{k}}$.
Of course, this is only a technical point, but it greatly facilitates the
derivation of the
functional-integral equation for the exact Greens function.
In momentum space the interaction part
of our hamiltonian can be written as
 \begin{eqnarray}
 {\cal{U}} & = & \frac{U}{N} \sum_{ {\bf{k}}_{1} \ldots {\bf{k}}_{4} }
 \delta^{\ast} ( {\bf{k}}_{1} + {\bf{k}}_{2} - {\bf{k}}_{3} - {\bf{k}}_{4} )
 \nonumber
 \\
& \times &
 \left[ c^{\dagger}_{{\bf{k}}_{1} \uparrow} c_{{\bf{k}}_{3} \uparrow}
	- \frac{1}{2} \delta_{{\bf{k}}_{1} , {\bf{k}}_{3}} \right]
 \left[ c^{\dagger}_{ {\bf{k}}_{2} + {\bf{\Pi}} \downarrow}
 c_{{\bf{k}}_{4} + {\bf{\Pi}}  \downarrow }
	- \frac{1}{2} \delta_{{\bf{k}}_{2} , {\bf{k}}_{4}} \right]
 \; \; \; ,
 \nonumber
 \\
 & &
 \label{eq:Hamiltonian3}
 \end{eqnarray}
where $\delta_{ {\bf{k}} , {\bf{k}}^{\prime} }$ is the usual
Kronecker-$\delta$, and
$\delta^{\ast} ( {\bf{k}} ) = \sum_{\bf{K}} \delta_{ {\bf{k}} , {\bf{K}} }$,
where
$\left\{ {\bf{K}} \right\}$ are the vectors of the reciprocal lattice. In
Eq.\ref{eq:Hamiltonian3} we have shifted the momentum of the last two operators
by ${\bf{\Pi}}$.
Obviously, this leaves $\delta^{\ast}$ invariant, so that the interaction term
can be expressed
entirely in terms of the components of the operator $C_{\bf{k}}$ defined in
Eq.\ref{eq:Cdef}.

We now introduce the imaginary-time $2 \times 2$ matrix Greens function
 \begin{equation}
 \underline{G} ( {\bf{k}} ,   \tau - \tau^{\prime} )
	=  - < {\cal{T}} \left[ C_{ {\bf{k}} } ( \tau )
 C^{\dagger}_{\bf{k}} ( \tau^{\prime} ) \right] >
 \; \; \; ,
 \label{eq:Gkdef}
 \end{equation}
where ${\cal{T}}$ denotes time ordering in imaginary time, and
the time evolution and thermal average are determined by ${\cal{H}}_{0}^{x} +
{\cal{U}}$.
The corresponding non-interacting Matsubara Greens function is
 \begin{eqnarray}
 \underline{G}^{(0)} ( {\bf{k}} , i\omega_{n} )
 & = &  \int_{0}^{1/T} d \tau e^{i \omega_{n} \tau}
 \underline{G}^{(0)} ( {\bf{k}} ,   \tau  )
 \nonumber
 \\
 & = & \left( \begin{array}{cc}
 i \omega_{n} - \epsilon_{\bf{k}} & h \\
 h & i \omega_{n} - \epsilon_{\bf{k} + \bf{\Pi}}
 \end{array} \right)^{-1}
 \; \; \; ,
 \label{eq:G0}
 \end{eqnarray}
where $\omega_{n} = \pi T ( 2n + 1 )$.
The self-energy matrix is defined as usual,
$\underline{\Sigma} ( {\bf{k}} , i \omega_{n} ) = \underline{G}^{(0) -1} (
{\bf{k}} , i
\omega_{n}) - \underline{G}^{-1} ( {\bf{k}} , i \omega_{n} )$.  The essential
simplification in
$d = \infty$  is that the momentum conservation can be
ignored\cite{MullerHartmann89}, and
we can replace $\delta^{\ast} ( {\bf{k}} ) \rightarrow 1/N$
the mathematical expressions for the Feynman diagrams.
The self-energy is then independent of ${\bf{k}}$,
and must be of the form
 \begin{equation}
 \lim_{d \rightarrow \infty} \underline{\Sigma} ( {\bf{k}} , i \omega_{n} )
\equiv
 \underline{\Sigma} ( i \omega_{n} )
 = \left( \begin{array}{cc}
 \Sigma ( i \omega_{n} )  & \Gamma ( i \omega_{n} ) \\
 \Gamma ( i \omega_{n} ) & \Sigma ( i \omega_{n} )
 \end{array} \right)
 \; \; \; .
 \label{eq:sigmadef}
 \end{equation}
Particle-hole symmetry implies that
 \begin{eqnarray}
 \Sigma (  i \omega_{n}  ) & = & - \Sigma ( - i \omega_{n} )
 \label{eq:sigmaodd}
 \\
 \Gamma (  i \omega_{n}  ) & =  & \Gamma ( - i \omega_{n} )
 \label{eq:gammaeven}
 \; \; \; .
 \end{eqnarray}
The exact on-site Greens function is then given by
 \begin{equation}
 \underline{G}_{n} = \frac{1}{N} \sum_{\bf{k}} \left[
 \underline{G}^{(0) -1} ( {\bf{k}} , i \omega_{n} ) - \underline{\Sigma}_{n}
\right]^{-1}
 \label{eq:Gndef}
 \; \; \; ,
 \end{equation}
where we use the abbreviation $\underline{G}_{n} = \underline{G} ( i \omega_{n}
)$ and
$\underline{\Sigma}_{n} = \underline{\Sigma} ( i \omega_{n})$.
{}From Eq.\ref{eq:G0} it is clear that in general the summand
on the left-hand side of Eq.\ref{eq:Gndef} depends on $\epsilon_{\bf{k}}$ and
$\epsilon_{\bf{k}+ \bf{\Pi}}$, and that therefore the summation cannot be
reduced to an integration over the density of states.
However, if we require that the non-interacting band structure satisfies the
perfect nesting
condition, Eq.\ref{eq:nesting}, the summand in Eq.\ref{eq:Gndef}
depends on $\epsilon_{\bf{k}}$ only, so that
 \begin{equation}
 \underline{G}_{n} = \int_{- \infty}^{\infty} d \epsilon \rho ( \epsilon )
\left(
 \begin{array}{cc}
 i \omega_{n} - \epsilon - \Sigma_{n} & h - \Gamma_{n} \\
 h - \Gamma_{n} & i \omega_{n} + \epsilon - \Sigma_{n}
 \end{array} \right)^{-1}
 \label{eq:Gn}
 \; \; \; .
 \end{equation}
The unique signature of perfect nesting is that the energy $\epsilon$
enters in the upper- and lower diagonal elements with opposite sign.

The functional-integral equation
for the exact Greens function is now derived in the standard
way\cite{MullerHartmann89}.
One defines a variational Greens function
 \begin{equation}
 \tilde{\underline{G}}_{n}^{-1} = \underline{G}_{n}^{-1} +
\underline{\Sigma}_{n}
 \; \; \; ,
 \label{eq:Gtildedef}
 \end{equation}
and a single-site impurity action
 \begin{eqnarray}
 S_{imp} & = &
 - \frac{1}{T} \sum_{n} C^{\dagger}_{n} \tilde{\underline{G}}_{n}^{-1}
 C_{n}
 \nonumber
 \\
 & + & U \int_{0}^{1/T} d \tau
 \left[ n_{\uparrow} ( \tau )  - \frac{1}{2} \right]
 \left[ n_{\downarrow} ( \tau ) - \frac{1}{2} \right]
 \; \; \; ,
 \label{eq:Simpdef}
 \end{eqnarray}
where $C^{\dagger} ( \tau ) = [ C_{ \uparrow}^{\dagger} ( \tau ) ,
 C_{\downarrow}^{\dagger} ( \tau ) ]$ are imaginary-time
two-component Grassmann fields, with Matsubara components
$C_{n} = T \int_{0}^{1/T} d \tau e^{ i \omega_{n} \tau}
C ( \tau )$,
and
$n_{\sigma} ( \tau ) =
C^{\dagger}_{\sigma} ( \tau ) C_{\sigma} ( \tau )$.
The functional-integral equation for the self-energies $\left\{
\underline{\Sigma}_{n} \right\}$
is then a $2 \times 2$ matrix equation
 \begin{eqnarray}
 \underline{G}_{n} & =  & - \frac{1}{T}
 \frac{ \int {\cal{D}} \left\{ C , C^{\dagger} \right\} \exp \left[ {- S_{imp}}
\right]
  C_{n}  C_{n}^{\dagger} }
 { \int {\cal{D}} \left\{ C , C^{\dagger} \right\} \exp \left[ {- S_{imp}}
\right] }
 \nonumber
 \\
 & \equiv &  - \frac{1}{T} < C_{n} C_{n}^{\dagger}  >_{S_{imp}}
 \; \; \; .
 \label{eq:func}
 \end{eqnarray}
Because $S_{imp}$ depends on all $\left\{ \underline{\Sigma}_{n} \right\}$, the
right-hand side  of Eq.\ref{eq:func} is in general a non-linear functional
of the self-energies, while the left-hand side
is a non-linear function of $\underline{\Sigma}_{n}$.
Hence,
Eq.\ref{eq:func} is a very complicated non-linear functional-integral equation.
To calculate $\Sigma_{n}$, one should first calculate
the exact Greens function of the  impurity model in Eq.\ref{eq:Simpdef}
for general choice of the $\left\{ \Sigma_{n} \right\}$, and obtain an explicit
expression for the right-hand side of Eq.\ref{eq:func}. After that, one should
solve
the resulting non-linear integral equation. Of course, such a calculation
can only be performed numerically. However, to examine the possibility
of symmetry breaking, it is not necessary the explicitly solve these equations.
%
%
\section{Staggered susceptibility and vertex function}
\setcounter{section}{3}
\setcounter{equation}{0}
\label{sec:proof1}

Suppose that $\underline{G}_{n} $ is a solution of Eq.\ref{eq:func}.
In the presence of a symmetry breaking field, the exact Greens function is of
the form
 \begin{equation}
 \underline{G}_{n} = \left( \begin{array}{cc}
 G_{n} & F_{n} \\
 F_{n} & G_{n}
 \end{array} \right)
 \; \; \; ,
 \label{eq:Gnmatrix}
 \end{equation}
where the anomalous Greens function $F_{n}$ is related
to the staggered magnetization $M (h,T)$
of the underlying Hubbard hamiltonian via
 \begin{equation}
 M(h,T) =  T \sum_{n} F_{n} = \sum_{n} < C^{\dagger}_{n} \sigma^{x} C_{n}
>_{S_{imp}}
 \; \; \; .
 \label{eq:Mh}
 \end{equation}
Note that {\it{antiferromagnetic}} symmetry breaking in the original Hubbard
model
translates into  {\it{ferromagnetic}} symmetry breaking in the impurity model.
Of course, the concept of antiferromagnetism is meaningless
in a world consisting of two degrees of freedom.
Because the
impurity model is essentially  zero-dimensional, no spontaneous symmetry
breaking
can occur in this model. Thus, the spin-susceptibility
$\tilde{\chi}$ of the impurity model, defined via
 \begin{eqnarray}
 \tilde{\chi}(T)  & = &
 T \int_{0}^{1/T} d \tau \int_{0}^{1/T} d \tau^{\prime}
 \nonumber
 \\
 & \times &
 <C^{\dagger} ( \tau ) \sigma^{x} C ( \tau )
 C^{\dagger} ( \tau^{\prime} ) \sigma^{x} C ( \tau^{\prime} )
 >_{S_{imp}}
 \label{eq:chin1}
 \; \; \; ,
 \end{eqnarray}
remains finite for all values of $T$ and $U$\cite{Schrieffer65}.
However, $\tilde{\chi}$ is not identical with the
staggered susceptibility $\chi$ of the Hubbard model, defined by
 \begin{equation}
 \chi (T) =  \lim_{h \rightarrow 0} \frac{ \partial M ( h,T ) }{\partial h }
 \label{eq:chidef}
 \; \; \; .
 \end{equation}
This is evident from the fact that the self-energies are complicated functions
of the external field, so that the derivative in Eq.\ref{eq:chidef}
does not simply produce the correlation function in Eq.\ref{eq:chin1}.
Below we shall make the relation between $\chi$ and $\tilde{\chi}$ precise,
and show that the
self-consistency condition, Eq.\ref{eq:func}, assures that $\chi$ can diverge
at
a finite temperature, while
$\tilde{\chi}$  remains finite.

To derive an expression for the staggered susceptibility
of the Hubbard model, let us first assume that
the hopping energies $t_{\bf{r}}$ in Eq.\ref{eq:Hamiltonian0}
are only non-vanishing for nearest neighbor sites.
In the weak coupling regime, the generalization to arbitrary hoppings,
subject to the restrictions mentioned earlier, is trivial and will be given
shortly.
Setting $t_{\bf{r}} = t / \sqrt{2 d}$ for ${\bf{r}}$ connecting neighboring
sites,
the density of states in $d = \infty$ is\cite{MullerHartmann89}
 \begin{equation}
 \rho ( \epsilon ) = \rho_{0} \exp \left[ - \pi \rho_{0}^2
 {\epsilon}^2 \right]
 \; \; \; ,
 \label{eq:rhogaus}
 \end{equation}
with $\rho_{0} \equiv \rho (0) = [ t \sqrt{2 \pi } ]^{-1}$.
The integration in Eq.\ref{eq:Gn} can then be done analytically, and we find
that the diagonal- and
off-diagonal elements of $\underline{G}_{n}$ are given by
 \begin{eqnarray}
 G_{n} & = & - \pi \rho_0  R_{n} \frac{ i \omega_{n} - \Sigma_{n} }{ \Omega_{n}
}
 \label{eq:Gnsol}
 \\
 F_{n} & = & \pi \rho_0 R_{n} \frac{ h - \Gamma_{n} }{ \Omega_{n} }
 \label{eq:Fnsol}
 \; \; \; ,
 \end{eqnarray}
where
 \begin{eqnarray}
 \Omega_{n} & = & \left[ - \left( i \omega_{n} - \Sigma_{n} \right)^2 +
 \left( h - \Gamma_{n} \right)^2 \right]^{1/2}
 \label{eq:omegandef}
 \\
 R_{n} & = & \mbox{erfc} \left[ \sqrt{\pi} \rho_{0} \Omega_{n}  \right]
 \exp \left[ \pi  \rho_{0}^2 \Omega_{n}^2  \right]
 \; \; \; .
 \label{eq:Rndef}
 \end{eqnarray}
Here $\mbox{erfc}[x]$ is the complimentary error function, and the root in
Eq.\ref{eq:omegandef} should be taken such that $\mbox{Re} \Omega_{n} \geq 0$.
The leading terms of $R_{n}$ for small and large $\rho_{0} |\Omega_{n}| $ is
 \begin{equation}
 R_{n} \sim \left\{
 \begin{array}{ll}
 1    & \mbox{for $  \rho_{0} |\Omega_{n} |  \ll 1$} \\
 \left[ \pi \rho_0 \Omega_{n} \right]^{-1} & \mbox{for $ \rho_{0} |\Omega_{n} |
\gg 1$}
 \end{array} \right.
 \label{eq:Rnlim}
 \; \; \; .
 \end{equation}
Thus, $R_{n}$ acts as a high-energy cutoff for frequency summations.
As long as the Matsubara sums
are dominated by the infrared regime $ \rho_{0} |\omega_{n} | \ll  1$, the
precise form of the
cutoff is irrelevant, so that we may set $R_{n} = 1$ , keeping in mind that all
frequency sums are cut off by an energy of the order of $\rho_{0}^{-1}$.

It is now also clear that inclusion of hoppings
between the sublattices
beyond the nearest neighbors will
lead to a different value of the cutoff $\rho_{0}$,
and a different functional form of the function $R_{n}$ at high frequencies.
However, the low-frequency behavior of
Eqs.\ref{eq:Gnsol}-\ref{eq:omegandef} will be unchanged, because the form of
these equations carries the unique signature of perfect nesting, and is
in this sense universal.
Hence, for $\rho_0 U \ll 1$
all calculations presented below hold also for arbitrary hoppings, provided we
substitute the value of $\rho_{0}$ and the cutoff function $R_{n}$
corresponding
to the particular choice of hoppings.

To examine the stability of the paramagnetic phase, we now calculate the
staggered susceptibility
$\chi$.
Substituting Eq.\ref{eq:Fnsol} into Eq.\ref{eq:Mh}, differentiating
with respect to $h$
and letting $h \rightarrow 0$, we obtain an exact relation between
the staggered susceptibility and the self-energy of the Hubbard model
 \begin{eqnarray}
 \chi & = &  T \sum_{n} \chi_{n}
 \label{eq:chisum2}
 \\
 \chi_{n} & = &  \lim_{h \rightarrow 0}
 \frac{ \partial F_{n}}{\partial h} =
 \pi \rho_0    R_{n} \frac{ \Lambda_{n} }{\Omega_{n}}
 \label{eq:chin2}
 \; \; \; ,
 \end{eqnarray}
where we have used the fact that in a parameter regime
in which the symmetry is not spontaneously
broken  $\lim_{h \rightarrow 0} \Gamma_{n}  =  0 $.
It is understood that
$\Omega_{n}$ and $R_{n}$ are now defined  by setting $h = \Gamma_{n} =
0$ in Eqs.\ref{eq:omegandef} and \ref{eq:Rndef}.
The vertex function $\Lambda_{n}$ is
defined via the Ward-identity
 \begin{equation}
 \Lambda_{n} = 1 - \lim_{h \rightarrow 0} \frac{ \partial \Gamma_{n} }{\partial
h}
 \; \; \; .
 \label{eq:vertexdef}
 \end{equation}
The divergence of the susceptibility is
controlled by the low-frequency behavior of the self-energies.
Recently it has been convincingly
demonstrated that in the weak coupling regime the
phase without broken symmetry is a Fermi
liquid\cite{Rozenberg92,Georges92a,Georges92b}.
Assuming that this is indeed correct, we now show that
for $\rho_{0} U  \ll 1$ the staggered susceptibility
diverges in the limit $T \rightarrow 0$.

Particle-hole symmetry
implies $\Sigma (0) = 0$, see Eq.\ref{eq:sigmaodd}.
The infrared behavior
of the sum in Eq.\ref{eq:chisum2} is then determined by the
finite temperature
effective mass- and vertex renormalization factors
$Z_{\omega} = m/m^{\ast} $ and $Z_{h}$,
 \begin{eqnarray}
 Z_{\omega}^{-1} & = &
1 - \frac{\mbox{Im} \Sigma_0 }{ \omega_0}
 \label{eq:Zmdef}
 \\
 Z_{h}^{-1} & = & 1 - \lim_{h \rightarrow 0} \frac{  \partial \Gamma_{0}  }{
\partial h}
 \label{eq:Zhdef}
 \; \; \; .
 \end{eqnarray}
Note that $Z_{h}^{-1}$
is precisely  the zero-frequency limit of the vertex defined in
Eq.\ref{eq:vertexdef}.
{}From \ref{eq:gammaeven} it is clear that $\Lambda ( i \omega_{n} )$ is an
even function of
frequency, so that the low-frequency behavior of Eq.\ref{eq:chisum2}
is determined by $\Lambda_{0}$.
We emphasize that in the strong-coupling regime
$ \rho_{0 } U \raisebox{-0.5ex}{$\; \stackrel{>}{\sim} \;$} 1$
the phase without broken symmetry is most likely a Mott insulator, where
Fermi liquid theory is not valid,
and the effective mass diverges\cite{Rozenberg92,Georges92a,Georges92b}.
Therefore our proof is necessarily restricted to the weak coupling regime.
Combining Eq.\ref{eq:Zmdef} with Eq.\ref{eq:omegandef},
we obtain for small frequencies
 \begin{equation}
 \Omega_{n} \sim
 Z_{\omega}^{-1}
 | \omega_{n}  |
 \; \; \; .
 \label{eq:omegasmall}
 \end{equation}
Substituting Eq.\ref{eq:omegasmall} into Eq.\ref{eq:chin2}, we see that
in the weak coupling regime the susceptibility is to leading logarithmic order
given by
 \begin{equation}
 \chi =  \rho_{0} \frac{Z_{\omega}}{Z_{h}}  \left[  L + O(1) \right]
 \; \; \; ,
 \label{eq:chinf}
 \end{equation}
where the dimensionless factor $L$ is
in the low temperature regime $\rho_{0} T \ll 1$ given by
 \begin{equation}
 L =  \pi T\sum_{n}  \frac{R_{n}}{|\omega_{n} |} =  \ln \left( \rho_{0} T
\right)^{-1}
 + O (1)
 \; \; \; .
 \label{eq:Ldef}
 \end{equation}
Note that $R_{n}$ merely provides a high-energy cutoff
of the order of $\rho_{0}^{-1}$ to the Matsubara sum,
the precise value of which is irrelevant.
Eq.\ref{eq:chinf} is completely general, although all interesting
physics is hidden in the renormalization factors $Z_{\omega}$ and $Z_{h}$.
Without further explicit knowledge of these factors,
there are a priori two possibilities: The first is that $Z_{\omega} / Z_{h}$
diverges
at a finite $T > 0$. Below
we shall show that in the regime  $0 < \rho_{0} U \ll 1$  this is indeed the
case,
because $Z_{h}$ vanishes while $Z_{\omega}$ remains finite.
The second possibility is that $ Z_{\omega} / Z_{h}$ remains finite.
It turns out that, at least in the weak coupling regime, this is only true for
$U =0$,
where $Z_{\omega} = Z_{h} = 1$. In this case the divergence of the
susceptibility
is due to the divergence of $L$
as $T \rightarrow 0$. In both cases, however, $\chi = \infty$ at $T=0$.
Note the essential role of particle-hole symmetry to assure
that $\Omega_{n}$ vanishes in the limit $\omega_{n} \rightarrow 0$, so that
the infrared cutoff for the sum in Eq.\ref{eq:Ldef} is
$\omega_{0} = \pi T$, and no other energy scale.
The only way in which the divergence could be avoided
is via a divergence
of the effective mass, so that $Z_{\omega}= 0$. This
possibility can be excluded, because the weak coupling phase without broken
symmetry is by assumption a Fermi liquid.
We therefore conclude that
the antiferromagnetic susceptibility is always infinite at $T=0$.
This is of course due to the nesting instability  built into our model.
Hence, the zero-temperature divergence of the staggered susceptibility survives
in the weakly interacting theory, and leads to an antiferromagnetic ground
state with
spontaneously broken symmetry.
This is the first main result of this work.
Of course, we cannot exclude the possibility that some other susceptibility
exhibits an even stronger divergence,
but this seems to be very unlikely.

%
%
\section{Ward-identity and proof of symmetry breaking for $T > 0$ at weak
coupling}
\setcounter{section}{4}
\setcounter{equation}{0}
\label{sec:proof2}

To proof that $T_N (U) > 0$, we now show that $Z_{h}$ vanishes for
$U > 0$ at a finite temperature.
To get some intuition how this might happen,
let us first calculate $Z_{h}$ within Hartree-Fock theory.
In this approximation  the self-energies are independent of frequency,
so that $\Sigma_{n} = 0$ by particle-hole symmetry, and $Z_{\omega} = 1$.
The off-diagonal self-energy is given by the first-order exchange diagram,
 \begin{equation}
 \Gamma^{(1)} = - U T \sum_{n} F^{(1)}_{n}
 \label{eq:HF}
 \; \; \; ,
 \end{equation}
where $F_{n}^{(1)}$ is obtained from our general expression
in Eq.\ref{eq:Fnsol} by setting $\Sigma_{n} = 0$ and $\Gamma_{n} =
\Gamma^{(1)}$.
Note that  Eq.\ref{eq:HF} is the usual
self-consistency equation for the mean-field gap $\Delta \equiv -
\Gamma^{(1)}$.
At low temperatures Eq.\ref{eq:HF} reduces to
 \begin{equation}
  \Gamma^{(1)} = \rho_{0} U  L [ \Gamma^{(1)} - h ]
  \label{eq:MFlow}
 \; \; \; ,
 \end{equation}
where the $L$ is defined in Eq.\ref{eq:Ldef}.
Differentiating both sides of Eq.\ref{eq:MFlow} with respect to $h$ and letting
$h \rightarrow 0$
yields
 \begin{equation}
 Z_{h}
 =   1 -  \rho_0 U   L
 \label{eq:lambdarpa}
 \; \; \; .
 \end{equation}
The mean-field estimate for the Neel temperature is then obtained from
$Z_{h} = 0$, which yields
 \begin{equation}
 T_N \propto \rho_{0}^{-1} \exp \left[ - (\rho_{0} U)^{-1}  \right]
 \label{eq:TNmf}
 \; \; \; .
 \end{equation}
{}From Eqs.\ref{eq:chisum2}, \ref{eq:chin2} and \ref{eq:lambdarpa}
it is also  obvious that the Hartree-Fock approximation for the self-energies
leads to the random-phase approximation for the susceptibility.

To show that Hartree-Fock theory is qualitatively correct, we now derive an
integral
equation for the vertex function $\Lambda_{n}$.
The crucial observation is that $Z_{h}$ can be determined by differentiation
of the self-consistency equation for the self-energies. The Hartree-Fock
self-consistency requirement, Eq.\ref{eq:HF}, is the simplest possible
approximation. In infinite dimensions we have an {\it{exact}}
self-consistency equation at our disposal.
The obvious procedure is then to
differentiate the off-diagonal
components of both sides of the functional-integral equation, Eq.\ref{eq:func},
with
respect to $h$, and taking then the limit $h \rightarrow 0$.
After a simple calculation we obtain in
this way an exact integral equation for the vertex function $\Lambda_{n}$
 \begin{equation}
 2 \pi  \rho_{0} R_{n} \frac{\Lambda_{n}}{\Omega_{n}} =
 \sum_{m} \tilde{\chi}_{n,m}
 \left[   1 +
 \Lambda_{m} \left(
 \frac{1}{\pi \rho_{0} R_{m} \Omega_{m}}   - 1 \right) \right]
 \; \; \; ,
 \label{eq:vertexint}
 \end{equation}
where the kernel $\tilde{\chi}_{n,m}$ is the
following correlation function of the impurity model,
 \begin{equation}
 \tilde{\chi}_{n,m}  =  \frac{1}{T^2}
 <C^{\dagger}_{n} \sigma^{x} C_{n} C^{\dagger}_{m} \sigma^{x} C_{m}
 >_{S_{imp}}
 \label{eq:kerneldef}
 \; \; \; .
 \end{equation}
Eq.\ref{eq:vertexint} is valid for all values of
$U$ and temperatures $T > T_{N} (U)$.
The kernel of this integral equation is given by the Matsubara components
$\tilde{\chi}_{n,m}$
of the susceptibility  of the impurity model.
The $\tilde{\chi}_{n,m}$ do not have a direct physical meaning, and should
be considered as auxiliary quantities.
The susceptibility of the impurity model, defined in Eq.\ref{eq:chin1},
can be written in a form analogous to Eqs.\ref{eq:chisum2} and \ref{eq:chin2},
 \begin{eqnarray}
 \tilde{\chi} & = & T \sum_{n} \tilde{\chi}_{n}
 \label{eq:chisum3}
 \\
 \tilde{\chi}_{n} & = & \sum_{m} \tilde{\chi}_{n,m} =
 \frac{1}{T^2} \sum_{m}
 <C^{\dagger}_{n} \sigma^{x} C_{n}
 C^{\dagger}_{m} \sigma^{x} C_{m}
 >_{S_{imp}}
 \; .
 \nonumber
 \\
 & &
 \label{eq:chitildedef}
 \end{eqnarray}
Of course, we do not have an analytic expression for $\tilde{\chi}_{n,m}$, but
for $\rho_{0}U \ll 1$ this kernel can in principle
be calculated perturbatively, and is free of singularities\cite{Zlatic83}.

To determine $Z_{h}$, we multiply both
sides of Eq.\ref{eq:vertexint} by $T$ and sum over $n$.
By construction,  summation of the
left-hand side of Eq.\ref{eq:vertexint} yields precisely
$2 \chi $, see Eqs.\ref{eq:chisum2} and \ref{eq:chin2}.
Recalling that $Z_{h}^{-1}$ is by definition the zero-frequency
limit of the vertex $\Lambda_{n}$, we obtain from Eq.\ref{eq:vertexint}
 \begin{equation}
 2 \chi  = \tilde{\chi} + Z_{h}^{-1}
 T \sum_{n} \tilde{\chi}_{n} \lambda_{n}
 \left[ \frac{1}{\pi \rho_{0} R_{n} \Omega_{n}}   - 1 \right]
 \; \; \; ,
 \label{eq:vertex2}
 \end{equation}
where
 \begin{equation}
 \lambda_{n} =  \frac{\Lambda_{n}}{\Lambda_{0}} =
 1 - \frac{\Lambda_{0} - \Lambda_{n}}{\Lambda_{0}}
 \label{eq:lambdadef}
 \; \; \; .
 \end{equation}
Re-arranging terms, \ref{eq:vertex2} can be written as
 \begin{equation}
 Z_{h} =
 Z - \tilde{\chi}^{-1} \left[   T \sum_{n}
 \frac{\lambda_{n}}{  \pi \rho_{0} R_{n} \Omega_{n} }
 { \tilde{\chi}_{n} } -
 2 {Z_{h}} \chi
 \right]
 \; \; \; ,
 \label{eq:vertex3}
 \end{equation}
with the numerical constant $Z$ given by
 \begin{equation}
 Z = \frac{ \sum_{n} \lambda_{n} \tilde{\chi}_{n}}
 { \sum_{n} \tilde{\chi}_{n} }
 \; \; \; .
 \label{eq:CCdef}
 \end{equation}
In the non-interacting limit a simple calculation
of the vertex function $\tilde{\chi}_{n}$ defined in Eq.\ref{eq:chitildedef}
gives
 \begin{equation}
 \tilde{\chi}_{n}^{(0)} = -2  G_{n}^{(0) 2} =
 2 \pi^2 \rho_{0}^{2} R_{n}^{(0 )2}
 \label{eq:tildechi0}
 \; \; \; ,
 \end{equation}
where $R_n^{(0)}$ is the value of $R_n$ at $U=0$, and
 \begin{equation}
 G_{n}^{(0)} = - i \pi \rho_{0} R_{n}^{(0)} \mbox{sign} \omega_{n}
 \label{eq:Gnzero}
 \end{equation}
is the non-interacting Greens function, see Eq.\ref{eq:Gnsol}.
Combining Eq.\ref{eq:tildechi0} with Eqs.\ref{eq:chisum2} and \ref{eq:chin2},
we obtain the exact identity
 \begin{equation}
 2  {Z_{h}} \chi = T \sum_{n}
 \frac{\lambda_{n} R_n }{ \pi \rho_{0}  \Omega_{n} }
 \frac{\tilde{\chi}_{n}^{(0)} }{R_n^{(0) 2}}
 \label{eq:chichi0}
 \; \; \; .
 \end{equation}
Substituting Eq.\ref{eq:chichi0} into Eq.\ref{eq:vertex3}, we
finally we arrive at
 \begin{equation}
 Z_{h} = Z -
 \frac{T}{\tilde{\chi}}
 \sum_{n} \frac{\lambda_{n} R_n }{ \pi \rho_{0}  \Omega_{n}}
 \left[ \frac{\tilde{\chi}_{n}}{R_n^2} -
 \left( \frac{\tilde{\chi}_{n}}{R_n^{ 2}} \right)_{U=0} \right]
 \; \; \; .
 \label{eq:vertex4}
 \end{equation}
This equation is the central result of this work.
It has the structure of a Ward-identity, relating
a vertex function (i.e. a derivative of a Greens function)
to the Greens function itself.
This identity is a direct consequency of the fact that
in infinite dimensions the exact local Greens function
satisfies a self-consistency condition.
Note that $Z_{h} = 1/ \Lambda_{0}$ appears also on the right hand side of
Eq.\ref{eq:vertex4}. However,
{\it{ $\Lambda_{0}$ appears exclusively through the regularized combination}}
$\lambda_{n} = 1 - ( \Lambda_0 - \Lambda_n)/ \Lambda_0 $,
which deviates from unity only due to the non-trivial
frequency dependence of the self-energies,
and remains manifestly finite even if $ \Lambda_{0}$ diverges.
{}From Eqs.\ref{eq:lambdadef} and \ref{eq:gammaeven} it is obvious that for
small $U$ and
$\omega_{n}$ we have
$\lambda_{n} = 1 + O (U^2 \omega_{n}^2)$.
Therefore all quantities on the right-hand side of Eq.\ref{eq:vertex4}
are free of singularities, and can be calculated perturbatively
if $\rho_{0} U \ll 1$.
To obtain the leading logarithmic divergence of the sum in Eq.\ref{eq:vertex4},
we can safely ignore vertex corrections and set $\lambda_{n} =1$.

{}From Eq.\ref{eq:vertex4} it is clear that $U = 0$
is a singular point in parameter space, because
only in this case the enumerator in the last term vanishes, so that
$Z_{h} = Z = 1$.
For any finite $U$, the singular frequency dependence
in the last term in Eq.\ref{eq:vertex4}
leads in the zero-temperature limit to an infrared divergence.

We emphasize that Eq.\ref{eq:vertex4} is exact
and valid for all values of $U$ and $T$, as long as the symmetry is
not spontaneously broken.
Note also the formal
similarity with the Hartree-Fock result.
In fact, if the right-hand side of Eq.\ref{eq:vertex4} is expanded to first
order
in $\rho_{0}U$, we recover Eq.\ref{eq:lambdarpa}.
In this approximation $\Lambda_{n}$ is independent of $n$, so that $\lambda_{n}
=1$.
{}From Eq.\ref{eq:CCdef} we see that this implies also $Z=1$.
Straight-forward perturbation theory gives to first order in $U$
 \begin{equation}
 \tilde{\chi}_{n} = \tilde{\chi}_{n}^{(0)} - U  G_{n}^{(0) 2}
 T \sum_{m} \tilde{\chi}_{m}^{(0)} + O (U^2)
 \label{eq:chiexpansion}
 \; \; \; .
 \end{equation}
Using then Eq.\ref{eq:Gnzero}, we see that
 \begin{equation}
 {\tilde{\chi}   }^{-1}
 \left[ \frac{\tilde{\chi}_{n}}{R_n^2} -
 \left( \frac{\tilde{\chi}_{n}}{R_n^{ 2}} \right)_{U=0} \right]
 = \pi^2 \rho_{0}^2   U + O (U^2)
 \; \; \; ,
 \label{eq:impHF}
 \end{equation}
and finally
 \begin{equation}
 Z_{h} = 1 - \rho_{0} U  \pi T \sum_{n} \frac{R_{n}}{| \omega_{n}|}
 \times \left[ 1 + O(\rho_0 U) \right]
 \; \; \; .
 \label{eq:HF2}
 \end{equation}
This is precisely the Hartree-Fock result,
which has been derived above in a much simpler way,
see Eq.\ref{eq:lambdarpa}.
However, Eq.\ref{eq:vertex4} is completely rigorous, and contains
all quantum fluctuations neglected in Hartree-Fock theory.

We now use the well known properties of the impurity model to show that
the Hartree-Fock value for the Neel temperature is asymptotically
exact in the limit $\rho_{0}U \rightarrow 0 $.
The right-hand side of Eq.\ref{eq:vertex4}
depends on the self-energies $\Sigma_{n}$,  the vertices $\lambda_{n} $, and
the
susceptibilities $\tilde{\chi}_{n}$ of the impurity model, which
by virtue of Eq.\ref{eq:func}
can also be considered as functionals
of the self-energies. The crucial point is now that
all these quantities can in principle be calculated perturbatively in powers
of $\rho_{0} U$, and that the perturbation theory has a finite radius of
convergence.
This follows from the facts (i) that
our model is by assumption a Fermi liquid, so that all interaction effects
can be absorbed in finite renormalization factors.
The second cornerstone of our proof is the fact (ii) that the Anderson
impurity model can be treated perturbatively in the weak coupling regime .
The lowest order correction
to the impurity susceptibility is given by Eq.\ref{eq:impHF}.
For impurity models with non-interacting Greens function of the form
$\tilde{G}_{n}^{(0)} = [ i \omega_{n} + i \Delta {\mbox{sign}} \omega_{n}
]^{-1}$
Zlati\'{c} and Horvati\'{c}\cite{Zlatic83} have proven that
the perturbation series for the spin susceptibility
is absolutely convergent for any $|U| < \infty$,
and that in the weak coupling regime the first few terms of the series yield
an extremely accurate approximation of the exact result.
Although the self-consistent Greens function $\tilde{G}_{n}$ in
Eq.\ref{eq:Simpdef} will
not be of the form assumed in Ref.\cite{Zlatic83}, it is extremely plausible
that the validity of
perturbation theory does not depend on the precise form
of the non-interacting Greens function\cite{Georges92b}.

Eqs.\ref{eq:lambdadef} and \ref{eq:CCdef} imply that
for small enough $U$ the constant $Z$ has a convergent expansion
with the first two terms given by
 \begin{equation}
 Z = 1 + z_{1} (\rho_{0} U)^2 + \ldots
 \; \; \; ,
 \label{eq:Csmall}
 \end{equation}
where the constant $z_{1}$ is of the order of unity.
Moreover, particle-hole symmetry guarantees that  the infrared cutoff
for the sum in Eq.\ref{eq:vertex4} is $\Omega_{0} = \pi T / Z_{\omega}$, so
that
the sum is logarithmically divergent in the limit $T \rightarrow 0$.
Hence, the last term in Eq.\ref{eq:vertex4} can be made arbitrarily large
by choosing the temperature small enough,
provided that $U > 0$ and the effective mass renormalization $Z_{\omega}$
remains finite.
Consequently, for arbitrarily small but positive $U$ there exists always
a temperature $T_{N} > 0$ such that $Z_{h}$ vanishes.

It follows by continuity that the Hartree-Fock result
becomes asymptotically exact for $\rho_{0}U \rightarrow 0$, and that therefore
Hartree-Fock theory correctly predicts a finite staggered magnetization
in the weak coupling regime.
The Neel temperature $T_{N} (U)$ can in principle be calculated
by expanding
$\Sigma_{n}$, $\lambda_{n}$, and $\tilde{\chi}_{n}$ to the desired order in
$\rho_{0} U$, and finding the temperature that satisfies
 \begin{equation}
 \sum_n \lambda_n \tilde{\chi}_n  =
 \sum_{n} \frac{\lambda_{n} R_n }{ \pi \rho_{0}  \Omega_{n}}
 \left[ \frac{\tilde{\chi}_{n}}{R_n^2} -
 \left( \frac{\tilde{\chi}_{n}}{R_n^{ 2}} \right)_{U=0} \right]
 \; \; \; .
 \label{eq:Neel}
 \end{equation}
Because for small $\rho_{0}U$ the solution of this equation
smoothly connects with the Hartree-Fock result, it has,
at least for small enough $\rho_{0}U$, a finite solution $T_{N} > 0$.
This completes our proof that
the Hubbard model in infinite dimensions has indeed a finite Neel temperature
in the weak coupling regime.
Combining Eqs.\ref{eq:chinf}
and \ref{eq:vertex4},
we obtain the following exact result for the staggered susceptibility
in the regime $ T , U \ll \rho_0^{-1}$,
 \begin{equation}
 \chi = \frac{ \rho_0 Z_{\omega}
 \ln \left( \rho_0 T \right)^{-1} }
 { Z -
 \frac{T}{\tilde{\chi}}
 \sum_{n} \frac{\lambda_{n} R_n }{ \pi \rho_{0}  \Omega_{n}}
 \left[ \frac{\tilde{\chi}_{n}}{R_n^2} -
 \left( \frac{\tilde{\chi}_{n}}{R_n^{ 2}} \right)_{U=0} \right] }
 \; \; \; .
 \label{eq:chistfinal}
 \end{equation}
Note that the form of this equation is very similar to
the standard result of the random-phase approximation.
In fact, to leading order in $\rho_0 U$
Eq.\ref{eq:chistfinal} reduces precisely to the random-phase
approximation for $\chi$.
%
%
%
\section{Conclusions}
\setcounter{section}{5}
\setcounter{equation}{0}
\label{sec:con}
The existence of an antiferromagnetic instability in the weak coupling regime
of the half filled Hubbard models with perfect nesting is not surprising.
Such an instability is predicted
by Hartree-Fock theory in all dimensions. While in one dimension this
instability is known
to be an artifact of mean-field theory, conventional wisdom is that at least in
high
enough dimensions Hartree-Fock theory becomes very accurate. In the present
work we have
used the machinery available in infinite dimensions to show that
for $\rho_0 U \ll 1$
Hartree-Fock theory and the random-pahse approximation are
very accurate in $d = \infty$.
We have shown that the
particle-hole symmetry of the half filled model, together with the perfect
nesting property of the
non-interacting energy dispersion, determine the behavior of the
susceptibility at low temperatures.
By differentiation of the exact self-consistency condition
available in $d= \infty $, we have obtained exact
Ward-identities between vertex functions and self-energies. The central
equations,
Eqs.\ref{eq:vertexint} and \ref{eq:vertex4}, hold for all values of $T$ and $U$
as long as the symmetry is not spontaneously broken,
and are the basis
for our proof of the antiferromagnetic instability in the weak coupling regime.
To show that the Neel temperature is indeed finite, we have assumed (i) that
the high-temperature phase without broken symmetry is a Fermi liquid.
For small enough $\rho_0 U$ this assumption can be justified by recent
analytical
and numerical work\cite{Rozenberg92,Georges92a,Georges92b}.
Another essential ingredient in our proof is (ii) that
in the weak coupling limit the susceptibility of the
impurity model can be calculated perturbatively and reduces to the
susceptibility of
the non-interacting theory in a continuous and smooth way.
Both assumptions, (i) and (ii), are on solid grounds\cite{Zlatic83} and can
hardly be
questioned. If the reader is willing to accept these assumptions, then our
proof
can be considered as rigorous.

We have not made any prediction about the strong coupling phase.
Although our fundamental equations are also valid in this case, the analysis
becomes more difficult,
because self-energies and impurity susceptibilities cannot be calculated
by straight-forward expansion in powers of $U$.
In particular, for
$ \rho_{0 } U \raisebox{-0.5ex}{$\; \stackrel{>}{\sim} \;$} 1$
we expect that
the phase without broken symmetry is a Mott
insulator, with diverging effective mass\cite{Rozenberg92,Georges92a}.
Thus, if the ground state
at strong coupling is an antiferromagnet (at least for nearest neighbor hopping
this is certainly the case), then we expect a direct transition between
a Mott insulator and an antiferromagnet as we lower the temperature.
The analysis of Eqs.\ref{eq:vertexint},
\ref{eq:vertex4}, and Eq.\ref{eq:Neel}, which are of course also valid at
strong coupling,
might reveal interesting new
phenomena, and is left for the future.

What is the relevance of our result to finite dimensions?
For $d < \infty$ the density of states $\rho ( \epsilon )$ vanishes outside a
fixed interval.
However, we have seen that in the weak coupling regime
the only parameter that determines the infrared behavior
of Matsubara sums is $\rho (0)$. Therefore, we believe that
in all dimensions $d \geq 3$ the Hartree-Fock theory becomes
asymptotically exact for $\rho_0 U \rightarrow 0$.
Obviously our result does not extrapolate to $d=1$.
But in this case the Hubbard model
can be solved exactly, and we know that
there is no spontaneous magnetization for all $U$, even at $T=0$\cite{Lieb68}.
In $d=2$ Hartree-Fock theory is completely
incorrect at $T > 0$,  because it predicts
spontaneous symmetry breaking at low temperatures, although
the rigorous Mermin-Wagner theorem\cite{Mermin66} tells
us that this can happen only at $T=0$.
Note also that in two dimensions $\rho (0) = \infty$
due to Van Hove singularities.
Thus, at weak coupling the extrapolation of the physics in $d = \infty$
to $d=2$ is not possible.
For nearest neighbor-hopping and large $U$,
the half filled square lattice Hubbard model is equivalent to
a two dimensional
quantum Heisenberg antiferromagnet, which seems to be ordered at $T=0$.
There remains the possibility that the order in the ground state
is destroyed in the weak coupling
regime\cite{Kopietz92a,Oitmaa91}.
It can be shown\cite{Kopietz92b} that even for arbitrarily small $U$ the
perturbation expansion is not governed by a small parameter, and
that, in contrast to $d \geq 3$, all "perturbative" corrections
to Hartree-Fock theory in $d=2$ are
of the relative order of unity.
Thus, $d=2$ seems to be closer to $d=1$, and a simple Hartree-Fock
description of the weak coupling regime seems not to be justified.

\begin{center}
{\small \bf {ACKNOWLEDGMENTS}}
\end{center}

I would like to thank
K. Sch\H{o}nhammer for several helpful discussions, and
P. G. van Dongen for instructive correspondence.

%

\end{document}